# Multi-Epoch Batch Least-Squares Positioning on the Lunar Surface Using LuGRE GPS L1 C/A Pseudorange Measurements

**Seunghwan Kim,** *University of Illinois Urbana-Champaign*
**Alex Minetto,** *Politecnico di Torino*
**Jacob Spagnolli,** *University of Colorado Boulder*
**Fabio Dovis,** *Politecnico di Torino*
**Dennis Akos,** *University of Colorado Boulder, Stanford University*

**Correspondence**
**Seunghwan Kim**
Department of Aerospace Engineering, University of Illinois Urbana-Champaign, Urbana, United States
sk232@illinois.edu

**Abstract**
The Lunar GNSS Receiver Experiment (LuGRE) mission captured Global Navigation Satellite System (GNSS) signals in the lunar region, yielding position, velocity, and timing estimates. Analysis of onboard lunar-surface position estimates shows kilometer-level positioning errors due to the long propagation distances, limited measurements, and unfavorable GNSS geometry. This paper presents time-multiplexed positioning using publicly available LuGRE L1 C/A pseudorange measurements. The proposed approach accumulates measurements by exploiting the receiver's stationary position in the Lunar-Centered Lunar-Fixed frame after landing. The effects of the number of accumulated epochs, receiver altitude constraint using coarse a-priori information, straight-line tangential altitude threshold, and minimum number of measurements required per epoch are examined. Compared with single-epoch least-squares positioning, this approach substantially improves positioning performance using single-band, single-constellation signals. With selected combinations of these parameters, batches of epochs randomly selected from multiple operations yield 100-m-level accuracy at the 67$^{th}$ percentile.

## 1. INTRODUCTION

With the growing interest in sustained human and robotic activities in the lunar region, positioning, navigation, and timing (PNT) services, which are indispensable for lunar exploration and operations, have drawn significant global attention. Under the broader LunaNet interoperability framework, dedicated lunar communication and PNT infrastructures are being developed, including NASA's Lunar Communication Relay and Navigation System (LCRNS), ESA's Lunar Communications and Navigation Services (LCNS), and JAXA's Lunar Navigation Satellite System (LNSS) (Giordano et al., 2022; Jaime Esper et al., 2025; Murata et al., 2023; National Aeronautics and Space Administration, 2025). Under its contract with NASA, Intuitive Machines has also outlined a phased plan for deploying its Lunar Data Network (LDN) satellites for lunar communication and PNT services. Beginning with the launch of the LDN-1 satellite in late 2026, Intuitive Machines plans to achieve full initial operating capability in 2029 (Mina et al., 2025). Advanced Space has also demonstrated crosslink-based autonomous cislunar navigation through the Cislunar Autonomous Positioning System Technology Operations and Navigation Experiment (CAPSTONE) mission (Thompson et al., 2022).

As part of these efforts, the possibility of leveraging well-established terrestrial Global Navigation Satellite System (GNSS) signals to support lunar PNT infrastructure has also been actively explored. Intuitive Machines' LDN-1, planned as the first satellite under NASA's LCRNS architecture, is expected to rely primarily on GNSS signals in combination with two-way radiometric tracking from LDN ground stations (Mina et al., 2025). Similarly, Japan's LNSS is expected to use GNSS signals to generate ephemerides for its navigation satellites (Murata et al., 2023). Alongside these implementation-oriented efforts, extensive theoretical and simulation studies have investigated the feasibility of using terrestrial GNSS signals for lunar PNT, including signal availability, link-budget performance, tracking feasibility, and achievable positioning accuracy in cislunar and lunar-surface environments. For example, Capuano et al. (2015) assessed Global Positioning System (GPS) and Galileo signal availability and code-based positioning performance along an Earth–Moon transfer trajectory up to the lunar surface. Similarly, Cheung et al. (2020) evaluated multi-constellation GNSS visibility and pseudorange-based positioning accuracy for a lunar vehicle in a near-rectilinear halo orbit using simulated GPS, Galileo, and GLONASS signals.

Building on these theoretical and simulation studies, the recent Lunar GNSS Receiver Experiment (LuGRE) provided the first known demonstration of terrestrial GNSS signal acquisition, tracking, and navigation near and on the Moon, using GPS L1 C/A and L5 signals and Galileo E1 OS and E5a signals during the Earth–Moon transit, lunar orbit, and lunar surface phases, with the payload data subsequently made publicly available. Parker et al. (2026) presented a comprehensive analysis of the mission observables, including the carrier-to-noise-density ratio (C/N0), signal availability, and receiver state estimates. Although their study provided the first comprehensive overview of the LuGRE flight results, the analysis was presented as an initial characterization of the mission data, with extended scientific analyses left for future work. Their signal-availability analysis highlighted the importance of multi-constellation and multi-frequency capability for lunar GNSS users: adding L5/E5a measurements to L1/E1 measurements improved the fraction of time with four or more simultaneously tracked distinct satellites, which the authors attributed to the higher transmit power and wider transmit-antenna beam of L5/E5a signals. Nevertheless, during lunar-surface operations, the availability of instantaneous four-satellite geometry remained limited, and single-constellation subsets did not provide sufficient simultaneous satellite visibility for standalone single-epoch PVT.

The onboard receiver produced single-point weighted least-squares (WLS) PVT solutions on the lunar surface when sufficient simultaneous measurements were available. However, the reported surface solutions were strongly limited by unfavorable geometry at lunar distances. From the lunar perspective, the GNSS satellites visible on the lunar surface are located on the far side of the Earth. As a result, only a small portion of the main-lobe power extending beyond the Earth, together with sidelobe signals, reaches the Moon, while the number of available measurements is further limited by occultation by the Earth. In a representative surface operation, range-position errors reached 1–100 km under poor geometry, whereas lateral-position errors remained below 1 km (Parker et al., 2026). These results indicate that GNSS-only PVT on the Earth-facing lunar surface is feasible but remains intermittent and geometry-limited, even when dual-constellation and dual-frequency measurements are used. This motivates the use of post-processing strategies that exploit the raw observables beyond the onboard single-epoch WLS framework.

In this work, we investigate a batch accumulated-pseudorange-based least squares approach using actual LuGRE lunar-surface data. Motivated by the fact that the LuGRE receiver remained stationary on the lunar surface, the proposed method accumulates pseudorange measurements across multiple epochs, rather than relying solely on simultaneous single-epoch visibility, to increase the number of usable observations and improve the conditioning of the lunar-surface positioning problem. Related multi-epoch batch least-squares approaches have previously been investigated for terrestrial GPS point positioning (Colestock, 1994). Furthermore, Vouch et al. (2025) investigated a regularized batch least-squares approach for cislunar kinematic positioning. In contrast, this study exploits the stationary position of the lunar receiver to accumulate GPS L1 C/A pseudorange measurements collected over multiple epochs for a common position estimate. A key advantage of this approach is that measurements collected during epochs that are individually insufficient for standalone PVT can still provide useful geometric information when combined with measurements from different epochs. This approach also provides geometric benefits by incorporating measurements collected under slightly different Earth-Moon viewing geometries. Although the same side of the Moon generally faces the Earth, lunar libration and related geometric variations, together with the motion of the GNSS satellites, can provide additional diversity in the satellite geometry as viewed by a lunar surface receiver. These advantages were confirmed through random epoch-selection tests, in which epochs were randomly selected and the pseudorange measurements contained in those epochs were accumulated to obtain position solutions through least squares estimation. The results show that this pseudorange-accumulation approach can provide lunar-surface position solutions with 67th-percentile errors on the order of 100 m using only L1 C/A pseudorange measurements. The detailed results are presented in Section 4.

To the best of the authors' knowledge, this study presents the first experimental demonstration of GNSS-only positioning on the lunar surface with sub-kilometer accuracy. This capability is also relevant to early lunar surface missions that may operate before dedicated lunar PNT systems reach full operational capability. During this transitional phase, only limited lunar communication and navigation infrastructure may be available. For example, ESA's Lunar Pathfinder is primarily designed to provide lunar communication relay services, and Coimbra et al. (2025) have investigated the opportunistic use of Doppler observables extracted from its communication downlink for PVT estimation of NASA's Endurance rover on the lunar surface. In that study, the proposed filter accumulated Doppler measurements over time and achieved convergence over several hours, with an a-priori position uncertainty specified as approximately 100 m. The present study addresses the complementary initialization problem by demonstrating that terrestrial GNSS pseudorange measurements can provide a lunar-surface position estimate at a comparable accuracy level, even when only single-constellation, single-frequency measurements are used. Thus, the proposed approach can serve as a practical initialization method for subsequent lunar-surface navigation filters that require an a-priori position estimate at this accuracy level.

This paper is organized as follows. Section 2 briefly introduces the LuGRE mission and publicly available datasets. Section 2.2 presents a brief analysis of LuGRE's onboard PVT solution performance. Section 3 describes the coordinate frame conversion used in this research to accumulate pseudorange measurements on the lunar surface. GNSS navigation messages provide the ephemeris parameters required to compute satellite positions and velocities in the Earth-centered Earth-fixed (ECEF) frame, whereas the stationary LuGRE receiver is fixed in the Lunar-centered Lunar-fixed (LCLF) frame. Therefore, accurate and consistent coordinate transformations are required to achieve positioning accuracy at the 100-m level. It should also be noted that, although the antenna used for the LuGRE mission was steerable to maintain Earth pointing, its phase-center position was assumed to remain fixed throughout this study because any displacement associated with antenna steering is negligible relative to the Earth-Moon distance and the positioning-error scale considered. Section 3.2 presents the theoretical formulation for lunar-surface position estimation using pseudoranges accumulated over time, together with an altitude-constrained approach. The proposed batch positioning technique assumes that the receiver position remains constant over the measurement accumulation interval, which is consistent with LuGRE's stationary operation on the lunar surface. Section 4 examines key factors affecting positioning performance, including the number of epochs accumulated, using positioning tests based on pseudoranges accumulated within individual OPs. These results confirm the benefits of multi-epoch accumulation for improving lunar-surface positioning accuracy. Additional random epoch-selection test results are also presented to corroborate the effects of these factors on positioning performance. Finally, Section 5 concludes the paper.

## 2. BACKGROUND

### 2.1. Introduction to LuGRE mission

The Lunar GNSS Receiver Experiment (LuGRE) launched aboard the Firefly Blue Ghost Mission 1 (BGM1) lander on January 15, 2025 as a joint NASA and Italian Space Agency (ASI) mission. The payload was developed by Qascom S.r.l. with the goal of extending GNSS-based positioning, navigation, and timing to the Moon. The LuGRE payload collected data in cislunar space as well as on the lunar surface in Mare Crisium. This mission expanded the proven Space Service Volume of GNSS, as it provided the first demonstration of GNSS use at distances exceeding 30 Earth radii and the first acquisition of GNSS signals on the lunar surface.

The LuGRE receiver was specifically designed to acquire and track GPS L1/L5 and Galileo E1/E5a signals as weak as 20 dB-Hz (Parker et al., 2026). In order to receive and decode such weak GNSS signals at lunar distances several important hardware and software considerations had to be made. Specialized hardware components included a High Gain Antenna (HGA) and multi-stage Low Noise Amplifier (LNA) (Dovis et al., 2023; Tedesco et al., 2023). The HGA consisted of a nine element passive planar antenna array from Haigh-Farr with a peak gain of 15.5 dBic and 14.9 dBic for L1 and L5 signals, respectively (Konitzer et al., 2024). This array remained pointed at Earth's center with a +/-1 degree pointing accuracy via 2 axis gimbaled support arm for the duration of data collection (Konitzer et al., 2024; Parker et al., 2022). The LNA was comprised of several stages featuring amplifiers, and dual band filtering via bandpass filters, and an equalizer (Dovis et al., 2023; Tedesco et al., 2023). The receiver was chosen to be a software defined radio (SDR), specifically an adapted version of the Qascom QN400-Space (Dovis et al., 2023; Tedesco et al., 2023). This allowed for real time processing (RTP) to compute onboard solutions as well as sample collection (SC) of raw in-phase/quadrature (I/Q) data that was downlinked for post-processing (Dovis et al., 2023). Software implementations included specialized processing of GNSS signals via Doppler compensation and high-sensitivity acquisition and tracking techniques (Dovis et al., 2022, 2023; Minetto et al., 2022). A post processing tool called the LuGRE quick Navigation Analysis and Reporting Tool (LuNART-q) was also developed to process both raw I/Q data and RTP data downlinked from the experiment (Minetto et al., 2025).

During the mission, data was collected at several different locations, which have been categorized into four stages. Commissioning (C) is data collected while the BGM1 lander was in Earth orbit. Transit (T) data consists of collections while the lander was enroute to the moon. Landing (L) and Surface (S) stages are data from when BGM1 was landing in and then stationary on the moon's Mare Crisium, respectively. Each of these stages provide unique scenarios for GNSS signal acquisition with things like differing dynamics, geometry, distances, the relative pointing of the GNSS transmitting and LuGRE receiving antenna patterns, and environmental conditions such as lunar dust. As such analyzing data from all mission stages is paramount to fully characterizing the use of GNSS signals on the moon.

The LuGRE mission dataset has been made publicly available by Parker et al. (2025), including raw GNSS observables for GPS L1, Galileo E1, GPS L5, and Galileo E5, as well as raw signal I/Q samples for the same frequency bands. Various Level 0 observables are provided, included in files named TLM_ACQ, TLM_NAV, TLM_RAW, TLM_CLK, TLM_EPH, IQS_L1, and IQS_L5. Although the dataset documentation provides detailed descriptions of the available data, a brief overview of files of interest in this research is provided below for readers' convenience. TLM files provide the telemetry data from the LuGRE receiver without any processing on the telemetry data. Specifically, TLM_NAV files include least-squares position, velocity, and time solutions, whereas TLM_RAW files include pseudorange, Doppler, carrier phase, and C/N0. Beyond the telemetry data, IQS files include raw in-phase and quadrature samples captured by the LuGRE receiver. The analysis of the IQS files can be found in Sciacca et al. (2026), where high-sensitivity acquisition enabled detection of GNSS signals in the L1/E1 and L5/E5a bands, including those from GPS, Galileo, Beidou, NavIC, and SBAS signals.

Among the measurements contained in the TLM_RAW files, this paper focuses on code pseudorange measurements collected by the LuGRE receiver while operating on the lunar surface. Additionally, Section 2.2 analyzes the accuracy of the onboard position solutions relative to the reference position using the TLM_NAV files. Analysis of both the TLM_RAW and TLM_NAV files reveals that GNSS signal availability is limited on the lunar surface and that the onboard position solutions achieve only kilometer-level accuracy. Specifically, the TLM_RAW measurements showed that no more than five distinct GNSS satellites were simultaneously available at any epoch when GPS and Galileo measurements in the L1/E1 and L5/E5a bands were considered. More than three measurements were available at only 6% of the epochs processed in real-time mode, implying that GNSS-based standalone single-epoch positioning could not be performed for most of the lunar-surface operation. Moreover, when only GPS L1 C/A measurements were considered, no standalone single-epoch positioning solution could be obtained at any epoch. Furthermore, analysis of the TLM_NAV files revealed that GNSS-only positioning on the lunar surface yielded a 67th-percentile 3D positioning error of 130 km relative to the reference landing position, with a median error of 65 km. Further details of this analysis are provided in Section 2.2.

### 2.2. Real-time processed PNT performance analysis

The positioning performance of the LuGRE onboard navigation solution was evaluated using the decoded TLM_NAV telemetry data. The analysis was restricted to the seven telemetry files corresponding to the lunar-surface operations (OPs 38_0, 40_0, 74_0, 76_0, 77_0, 77_1, 78_1). Each navigation record contains the receiver time, number of satellites used by the navigation solution, receiver's position, velocity, and clock estimates, and dilution of precision metrics. For the present analysis, the three-dimensional position solution and the timestamp associated with each navigation record were extracted.

Since the receiver timestamps are provided in GPS time, these receiver times were first converted to UTC by considering the leap seconds and subsequently converted to ephemeris time (ET) using NASA's Spacecraft, Planet, Instrument, C-matrix, Events (SPICE) function *cspice_str2et*. Each onboard position solution expressed in ECEF frame was then transformed into the LCLF Mean Earth/Polar Axis (ME) frame. The detailed coordinate transformation procedure, including the ECEF-to-LCLF transformation, the translation of the origin from the Earth center to the Moon center, and the transformation into the lunar ME frame, is described in Section 3.

The Blue Ghost landing-site coordinates reported by the Lunar Reconnaissance Orbiter Camera (LROC) were used as the reference position (Robinson, 2025a). The landing site was defined by a latitude and longitude of $18.5623°\mathrm{N}, 61.8103°\mathrm{E}$, and an altitude of −3,650 m. To obtain the corresponding Cartesian reference position, the mean lunar radius was calculated from the three lunar radii provided in the SPICE planetary constants kernel. The reported altitude was then added to this mean radius to determine the radial distance of the landing site from the lunar center. The reference position was converted from latitude, longitude, and radius to lunar body-fixed Cartesian coordinates using the SPICE *cspice_latrec* function. The 3D cartesian errors were computed as the Euclidean separation between the onboard position estimates and the reference landing-site position. The error was calculated for every decoded navigation solution contained in the lunar-surface telemetry files.

The resulting error distribution was characterized using several descriptive statistics. The mean, median errors, and the 5th, 67th, and 95th percentiles were obtained as 3700 km, 65 km, 4 km, 130 km, 9600 km, respectively, and these values are used as a baseline for comparison with the proposed accumulated multi-epoch positioning method. It should be noted that the TLM_NAV data, which provide 3D position estimates in the Earth-centered Earth-fixed (ECEF) frame, were used for this analysis. These positions were transformed into the lunar ME frame using the coordinate transformation described in Section 3, and compared with the reference position, which is also expressed in the ME frame.

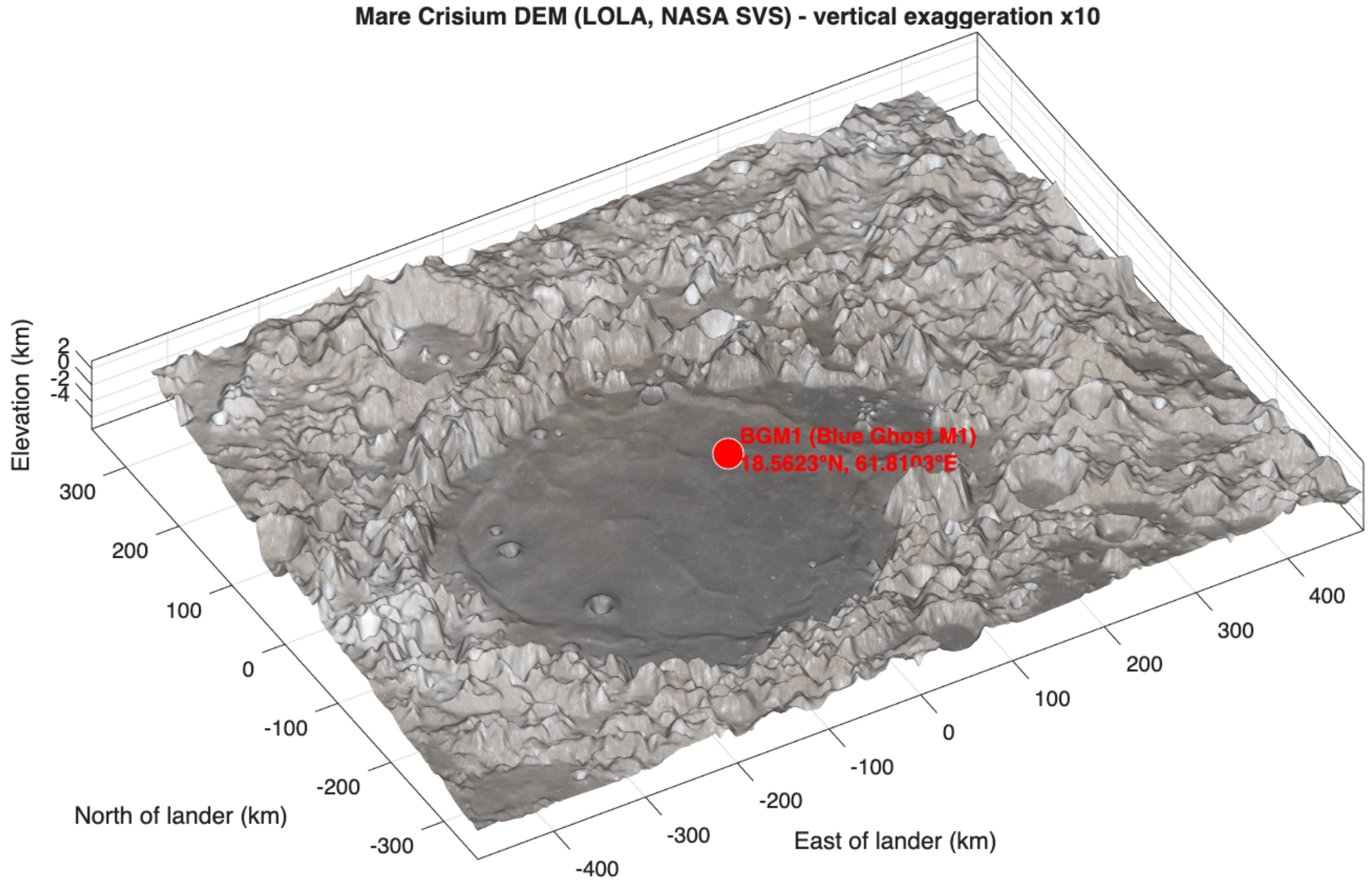

**Figure 1 Three-dimensional visualization of the lunar terrain surrounding the Blue Ghost Mission 1 (BGM1) landing site in Mare Crisium, generated using the Lunar Orbiter Laser Altimeter (LOLA) digital elevation model (NASA's Scientific Visualization Studio, 2019). The red marker indicates the reference landing position (18.5623°N, 61.8103°E). The terrain elevation is vertically exaggerated by a factor of 10.**

## 3. METHODOLOGY

### 3.1. Coordinate frames and transformations for lunar-surface positioning

The LuGRE receiver is stationary in the LCLF frame, whereas the GNSS satellite positions obtained from the broadcast ephemerides are expressed in an ECEF frame (Anthony, 2022). Therefore, the receiver and satellite positions must be expressed in a common coordinate frame to compute the geometric ranges between the receiver and the GNSS satellites. Accordingly, NASA's SPICE toolkit was used to transform the broadcast GNSS satellite positions into the LCLF frame in which the receiver position was estimated. The GPS broadcast ephemerides used in this study were obtained from NASA's Crustal Dynamics Data Information System (CDDIS), which also served as the source of GPS broadcast ephemerides provided to the LuGRE onboard receiver as aiding information for real-time navigation (Parker et al., 2026) .

Three principal coordinate frames were used in the transformation. First, the GNSS satellite position computed from the broadcast ephemeris was treated as an ECEF position at the signal transmit time. In the SPICE implementation, the ITRF93 frame was used as the practical ECEF frame. Second, the J2000 inertial frame was used as an intermediate frame in which the positions of the Earth, Moon, and GNSS satellite could be combined consistently. Third, the lunar ME frame was implemented using the SPICE MOON_ME frame. The receiver position remained fixed in this frame throughout the accumulated lunar-surface observations.

The calculation of geometric ranges between the GNSS satellites and the receiver, which entailed coordinate frame conversion, was performed using the following procedures. Let $t_{tx,k,j}$ and $t_{rx,k}$ denote the signal transmit time and receiver receive time, respectively, for the $j$-th satellite measurement at epoch k. The position of the satellite relative to the Earth, obtained from the broadcast ephemeris, is denoted by $\vec{r}_{s/E}^{ECEF(t_{tx,k,j})}$, where the superscript $ECEF$ denotes the ECEF frame at time $t_{tx,k,j}$ and the subscript $s/E$ indicates the position of the satellite relative to the Earth center. It should be noted that the GPS transmit and receive time tags were converted to Ephemeris Time (ET) before evaluating the SPICE ephemerides and frame transformations. Within the SPICE system, ET is a numerical representation of Barycentric Dynamical Time (TDB), expressed in seconds past the J2000 epoch (Park et al., 2021). The GPS time tags were first converted to UTC by applying the GPS-UTC leap seconds, after which the resulting UTC epochs were converted to ET using *cspice_str2et*.

The satellite position represented in ECEF frame is first transformed into the J2000 inertial frame at the signal transmit time:

$$\vec{r}_{s/E}^{J2K(t_{tx,k,j})} = {}^{J2K}\boldsymbol{T}_{ECEF}\left(t_{tx,k,j}\right) \times \vec{r}_{s/E}^{ECEF(t_{tx,k,j})} \tag{1}$$

where ${}^{J2K}\boldsymbol{T}_{ECEF}\left(t_{tx,k,j}\right)$ is the direction cosine matrix from ITRF93 to J2000 evaluated at the transmit time.

The Moon's position with respect to the Earth at $t_{rx,k}$ in the J2000 frame is then obtained through *cspice_spkpos* function provided by the SPICE toolkit. This is to form the GNSS satellite's position at $t_{tx,k,j}$ from the Moon's origin at $t_{rx,k}$ as follows:

$$\vec{r}_{s/M}^{J2K} = \vec{r}_{s/E}^{J2K(t_{tx,k,j})} - \vec{r}_{M/E}^{J2K(t_{rx},k)} \tag{2}$$

This expression accounts for the different origins of the broadcast satellite coordinates and the lunar receiver coordinates. In particular, the Earth and satellite positions are evaluated at the transmit time, whereas the Moon-center position is evaluated at the receive time.

The Moon-centered J2000 position was subsequently rotated into the lunar-fixed frame at the receive time:

$$\vec{r}_{s/M}^{ME} = {}^{ME}\boldsymbol{T}_{J2K}\left(t_{rx,k}\right) \times \vec{r}_{s/M}^{J2K} \tag{3}$$

where $ME$ denotes the lunar ME frame and ${}^{ME}\boldsymbol{T}_{J2K}\left(t_{rx,k}\right)$ is the J2000-to-ME direction cosine matrix evaluated at the signal reception time. This rotation was obtained using *cspice_pxform* function.

Combining the preceding operations, the complete coordinate transformation can be written as:

$$\vec{r}_{s/M}^{ME} = {}^{ME}\boldsymbol{T}_{J2K}\left(t_{rx,k}\right) \times \left[{}^{J2K}\boldsymbol{T}_{ECEF}\left(t_{tx,k,j}\right) \times \vec{r}_{s/E}^{ECEF(t_{tx,k,j})} - \vec{r}_{M/E}^{J2K(t_{rx},k)}\right] \tag{4}$$

The receiver position, $\vec{r}_r^{ME} = [x, y, z]^T$, was directly estimated in the same ME frame. Therefore, the modeled geometric range for the $j$-th measurement at epoch $k$ was computed as:

$$\hat{\rho}_{k,j} = \left|\vec{r}_r^{ME} - r_{s/M}^{ME}\right| \tag{5}$$

The procedure described above corresponds to the baseline geometric range model used in this study. It provides a consistent transformation of broadcast Earth-fixed GNSS satellite positions and the lunar-fixed receiver position into a common frame while accounting for the Earth-Moon origin displacement and the frame orientations at the relevant epochs.

The Earth–Moon ephemeris and the orientation of the lunar ME frame were obtained from the NASA Jet Propulsion Laboratory's Development Ephemeris 440 (DE440) and the associated SPICE kernels. According to Park et al. (2021), in DE440, numerically integrated orbits were fitted to observations made from the ground and in space. The Moon's orbit was determined primarily from lunar laser ranging (LLR) observations, while the directions of the PA axes were taken from analyses of Gravity Recovery and Interior Laboratory (GRAIL) data. The lunar libration angles defining the time-dependent orientation of the PA frame were integrated simultaneously with the orbital motion. The root-mean-square of the recent LLR range residuals was approximately 1.3 cm.

In this study, the DE440-based ME frame realization, MOON_ME_DE440_ME421, was used. This frame is defined by a constant rotational offset from the DE440 PA frame, MOON_PA_DE440, such that it is closely aligned with the ME frame associated with DE421. The time-dependent orientation of MOON_PA_DE440 relative to the International Celestial Reference Frame (ICRF) was obtained from the DE440 Planetary Constants Kernel, moon_pa_de440_200625.bpc. The parameters of the constant rotation are specified in the accompanying frame kernel, moon_de440_250416.tf, and were derived by comparing lunar retroreflector coordinates expressed in the DE440 PA frame with the corresponding coordinates in the DE421 ME frame (Bachman, 2025; Park et al., 2021).

The LuGRE lunar-surface operations analyzed in this study span approximately two weeks, whereas the receiver position was assumed to be stationary with respect to the lunar ME frame (MOON_ME_DE440_ME421). In principle, time-dependent errors in the lunar ephemeris and libration model could cause a fixed lunar-surface point to exhibit a small apparent epoch-to-epoch displacement in the ME frame, thereby introducing an inconsistency in the stationary-receiver assumption. Although a formal accuracy estimate for MOON_ME_DE440_ME421 is not publicly available to the best of the authors' knowledge, several studies on lunar ephemerides suggest that the error scale associated with ephemeris and frame-realization differences is significantly smaller than the 100-m-level positioning-error scale addressed in this study. Rambaux et al. (2026) compared the DE440 PA and the Integration Numérique de l'Observatoire de Paris 21a (INPOP21a) PA realizations and found that the lunar retroreflector coordinates differed by up to approximately 1 m between the two PA-frame realizations. In addition, the maximum sampled orientation difference between the DE421 ME frame (MOON_ME_DE421) and the DE440-based ME frame (MOON_ME_DE440_ME421) over the 2000-2040 period, an interval that includes the March 2025 LuGRE observations, corresponded to approximately 53.4 cm of great-circle displacement at the lunar surface (Bachman, 2025). Although these comparisons do not provide a formal bound on the apparent receiver displacement over the two-week observation span, they indicate that the examined inter-realization differences are at the submeter-to-meter scale. On this order-of-magnitude basis, any resulting inconsistency in the stationary-receiver assumption is unlikely to be a dominant contributor to the 100-m-level positioning errors reported in this study.

## 3.2. Time-multiplexed pseudorange-based batch least squares method

This section describes the batch least-squares formulation used for lunar-surface positioning with accumulated pseudorange measurements. Because the LuGRE receiver is stationary on the lunar surface with respect to LCLF frame during the accumulated observations, the receiver position is assumed to be common across all selected epochs. In contrast, the receiver clock bias is allowed to vary from epoch to epoch. The state vector is therefore defined as

**State vector:**

$$\mathbf{X} = \left[\vec{\mathbf{r}}_{\mathrm{ME}}^{\mathrm{T}}, \mathrm{b}_1, \cdots, \mathrm{b}_\mathrm{k}, \cdots, \mathrm{b}_\mathrm{n}\right]^{\mathrm{T}} \tag{6}$$

where $\vec{\mathbf{r}}_{ME} = [x, y, z]^T$ is the three-dimensional position of the LuGRE antenna in the ME frame, $\mathrm{b}_\mathrm{k}$ is the receiver clock bias at the $\mathrm{k}$-th accumulated epoch expressed in meters, and $\mathrm{n}$ is the total number of accumulated epochs. For example, when one epoch is selected from each of the seven lunar-surface OPs, the state vector contains 7 epoch-wise receiver clock biases in addition to the three common position components, yielding 10 state variables in total.

Measurement vector is formed as

**Measurements:**

$$\mathbf{Y} = \left[\rho_{1:\mathrm{i}_1}^{(1)}, \rho_{1:\mathrm{i}_2}^{(2)}, \cdots, \rho_{1:\mathrm{i}_\mathrm{n}}^{(\mathrm{n})}, \mathrm{h}_{\mathrm{priori}}\right] \tag{7}$$

where $\rho_{1:\mathrm{i}_\mathrm{k}}^{(\mathrm{k})}$ denotes the group of $\mathrm{i}_\mathrm{k}$ pseudoranges included in the $\mathrm{k}$-th epoch, and $\mathrm{h}_{\mathrm{priori}}$ is the a-priori altitude information used as an additional constraint. Since this study also examines the efficacy of altitude-constraint on positioning performance, the a-priori receiver's altitude information is included in the measurement vectors, as a pseudo-measurement. From the estimated x, y, z coordinates of the receiver, the predicted altitude is derived by subtracting the Moon's radius from the Euclidean norm of the estimated position. The difference between the a priori altitude and the predicted altitude is then included in the WLS residual vector and weighted according to the assumed altitude uncertainty.

Each pseudorange comprising the measurement vector can be modeled as

$$\rho_{\mathrm{k,j}} = \left|\vec{r}_r^{LCLF} - \vec{r}_{s/M}^{LCLF}\right| + b_k - B_{k,j} + A_{k,j} + \epsilon_{k,j} \tag{8}$$

where $b$ and $B$ denote clock biases of the receiver and satellites, respectively, $A$ is the atmospheric delay that includes ionospheric, tropospheric, and plasmaspheric impacts, and $\epsilon$ is measurement noise, which is assumed to be white and Gaussian in this paper. The satellite clock bias can be corrected using the clock-correction parameters contained in the broadcast navigation messages. The unprocessed broadcast navigation messages obtained from NASA CDDIS were used in this research.

It should also be noted that because no dedicated ionospheric/plasmaspheric correction model is applied in this study, unmodeled propagation delays remain in the single-frequency pseudorange measurements. In the proposed multi-epoch formulation, a separate receiver clock bias is estimated for each accumulated epoch. Consequently, the epoch-common component of the unmodeled propagation delay can be absorbed into the corresponding receiver clock-bias state, together with the true receiver clock offset and other common-mode range errors.

Therefore, the estimated epoch-wise clock biases should be interpreted as common range biases rather than pure receiver timing estimates.

Instead of estimating the ionospheric/plasmaspheric group delay at each epoch, this study applies a straight-line tangential altitude (SLTA)-based screening criterion to reduce the impact of low-SLTA propagation effects. SLTA is defined as the minimum altitude of the straight-line propagation path between a GNSS satellite and the lunar receiver above the Earth's surface. Specifically, epochs containing any measurements with SLTA below the prescribed threshold are excluded from the accumulated least-squares system. In this study, SLTA thresholds of 0, 500, and 1,000 km are considered to evaluate the effects of this screening criterion on positioning performance. This screening does not remove all unmodeled propagation delays, but it reduces the likelihood that large low-SLTA delay errors contaminate the epoch-wise clock-bias estimates or project into the common position estimate. Since conventional single-frequency ionospheric correction models, including the Klobuchar model, were developed primarily for terrestrial GNSS users, they are not directly applicable to lunar GNSS observations without modification due to differences in signal propagation geometry (Iiyama & Gao, 2025; Klobuchar & Member, 1987; Montenbruck & González Rodríguez, 2020). Dual-frequency ionosphere-free combinations of L1 and L5 pseudorange measurements could also be considered to mitigate ionospheric delays. However, given the relatively low carrier-to-noise-density ratios of GNSS signals at lunar distances, the noise amplification associated with such combinations may offset the benefits of ionospheric delay mitigation. Furthermore, dual-frequency combinations may reduce measurement availability, as not all GNSS satellites transmit signals on both frequency bands, and simultaneous reception at lunar distances may be further limited by weak signals and differences in satellite antenna patterns between frequency bands (Vouch et al., 2025). Therefore, this study applies SLTA-based screening to reduce the impact of potentially large propagation delays using single-frequency measurements. The resulting positioning performance is presented in Section 4.2.2.

The position-geometry matrix corresponding to epoch ($k$) is defined as

$$\boldsymbol{G_k} = [g_1 \quad g_2 \quad \cdots \quad g_{i_k}]^T \tag{9}$$

where $g_k = \left.\frac{\partial \rho_k}{\partial \vec{r}}\right|_{\hat{\vec{r}}} = \frac{\hat{\vec{r}} - \vec{r}_s}{\rho}$ is the unit vector from a GNSS satellite to the estimated receiver position at epoch $k$, and $i_k$ is the number of pseudorange measurements available at epoch $k$.

Accordingly, the Jacobian matrix of the accumulated least-squares system is expressed in block form as

$$\boldsymbol{H} = \begin{bmatrix} \boldsymbol{G_1} & \mathbf{1}_{\boldsymbol{i_1}\times\mathbf{1}} & \mathbf{0} & \cdots & \mathbf{0} \\ \boldsymbol{G_2} & \mathbf{0} & \mathbf{1}_{\boldsymbol{i_2}\times\mathbf{1}} & \cdots & \mathbf{0} \\ \vdots & \vdots & \vdots & \ddots & \vdots \\ \boldsymbol{G_n} & \mathbf{0} & \mathbf{0} & \cdots & \mathbf{1}_{\boldsymbol{i_n}\times\mathbf{1}} \\ \vec{u}_r^T & 0 & 0 & \cdots & 0 \end{bmatrix} \tag{10}$$

where $\vec{u}_r^T = \left.\frac{\partial h}{\partial \vec{r}}\right|_{\hat{\vec{r}}} = \frac{\hat{\vec{r}}^T}{\|\hat{\vec{r}}\|} = \begin{bmatrix} \frac{\hat{x}}{\|\hat{\vec{r}}\|} & \frac{\hat{y}}{\|\hat{\vec{r}}\|} & \frac{\hat{z}}{\|\hat{\vec{r}}\|} \end{bmatrix}$ is the outward radial unit vector from the lunar center toward the estimated receiver position. The vector $\mathbf{1}_{\boldsymbol{i_n}\times\mathbf{1}}$ is a column vector of ones, where $\boldsymbol{i_n}$ is again the number of pseudorange measurements available at epoch $n$.

Under this formulation, the nonlinear weighted least-squares (WLS) problem is solved iteratively, starting from an initial state estimate of $\hat{\mathbf{x}}^{(1)} = \left[\mathrm{R}_{\mathrm{moon}}, 0,0, \mathbf{0}_{\mathbf{1}\times\mathbf{n}_{\mathbf{epoch}}}\right]^{\mathrm{T}}$ used in this research, where $\mathrm{R}_{\mathrm{moon}}$ is the radius of the Moon, and $\mathbf{0}_{\mathbf{1}\times\mathbf{n}_{\mathbf{epoch}}}$ is an $\mathrm{n}_{\mathrm{epoch}}$ dimensional zero vector representing the initial receiver clock biases for the $\mathrm{n}_{\mathrm{epoch}}$ accumulated epochs. At each iteration $q$, the state is updated as follows:

$$\Delta\hat{\mathbf{x}}^{(\mathrm{q})} = \left(\mathbf{H}^{\mathrm{T}}\mathbf{W}\mathbf{H}\right)^{-1}\mathbf{H}^{\mathrm{T}}\mathbf{W}\left(\mathbf{y} - \hat{\boldsymbol{y}}^{(\boldsymbol{q})}\right), \qquad \hat{\mathbf{x}}^{(\mathrm{q}+1)} = \hat{\mathbf{x}}^{(\mathrm{q})} + \Delta\hat{\mathbf{x}}^{(\mathrm{q})} \tag{11}$$

where $\hat{\mathbf{x}}^{(\mathrm{q})}$ is the best state estimate at iteration $q$, $\hat{\boldsymbol{y}}^{(\boldsymbol{q})}$ is a predicted measurement vector evaluated at $\hat{\mathbf{x}}^{(\mathrm{q})}$, $\mathbf{W} = \mathrm{diag}\left(\sigma_\rho, \ldots, \sigma_\rho, \sigma_{\mathrm{H}}^{-2}\right) = \mathrm{diag}\left(1, \ldots, 1, \sigma_{\mathrm{H}}^{-2}\right)$ is the weighting matrix that assigns unit weight to each pseudorange measurement and a weight of $\sigma_{\mathrm{H}}^{-2}$ to the altitude pseudo-measurement This weighting accounts for the uncertainty in the a priori altitude information and allows its impact on positioning performance to be evaluated, as will be discussed in Section 4.2.1.

This formulation is referred to here as time-multiplexed pseudorange accumulation because pseudoranges obtained at different times are combined to estimate a single fixed receiver position, while a separate receiver clock bias is estimated for each epoch. Because a new receiver-clock state is introduced for every accumulated epoch, an epoch with only one pseudorange measurement provides little independent information about the common receiver position, whereas epochs containing multiple measurement can provide additional geometric information. Therefore, the positioning contribution of an epoch depends not only on the number of accumulated epochs but also on the number of simultaneous measurements available within each epoch.

Although the present study uses batch least-squares because all measurements are processed on the ground, the same formulation can be implemented recursively when computational resources are limited.

## 4. RESULTS

This section presents the single-OP positioning results to show the decrease in 3D positioning error as more epochs are accumulated and to identify the key factors affecting positioning performance on the lunar surface. It also presents the results of random epoch-selection tests conducted to characterize the statistical performance of GNSS-only positioning on the lunar surface using measurements accumulated over time. The effects of various settings, including the number of epochs accumulated from each operation (OP), the use of an altitude constraint, SLTA threshold values used to exclude epochs containing any measurements with SLTA values below the threshold, and the minimum number of measurements required per epoch, were analyzed to evaluate their effects on positioning performance. Epochs were randomly selected from each LuGRE OP, and 3,000 randomized trials were performed unless otherwise noted. This following results focus on the 67th and 95th percentiles.

### 4.1. Single-Operation Analysis of Multi-Epoch Positioning Performance

To examine the effect of measurement accumulation over time on 3D positioning error, measurements within a single operation were accumulated over time for batch positioning. The positioning performance generally followed the quality of the GNSS satellite geometry, which was quantified by the PDOP under the assumption of perfect knowledge of the receiver clock bias (hereafter referred to as known-clock PDOP), defined as

$$PDOP_{perfect} = \sqrt{trace(\boldsymbol{Q}_{perfect})}$$

where $\mathbf{Q}_{perfect} = (\boldsymbol{S}^T\mathbf{S})^{-1}, \boldsymbol{S} = [\boldsymbol{G_1} \quad \boldsymbol{G_2} \quad \cdots \quad \boldsymbol{G_n}]^T$

Figure 2 shows an example of the relationship among the number of accumulated measurements, the 3D positioning error, and the known-clock PDOP for OP 76_0. Epochs containing three pseudorange measurements, the maximum number available in this OP, were selected and accumulated over a period of up to five minutes. The known-clock PDOP decreased as the 3D positioning error decreased over time. In this case, the known-clock PDOP decreased from approximately 46 to 4 after five minutes of accumulation, although the rate of decrease gradually diminished as more epochs were accumulated. Similarly, the 3D positioning error decreased from approximately 70,000 km, falling sharply at first and then more gradually, to 21 km after five minutes of accumulation. The least-squares estimation system became overdetermined after two epochs were accumulated, yielding six measurements for five unknown state variables: three Cartesian position components and two epoch-specific receiver clock biases. By the end of the five-minute accumulation period, a total of 900 measurements had been accumulated and 303 state variables were estimated.

The altitude constraint, enabled by a priori knowledge of the lunar-surface height, significantly improved positioning performance, as shown in Figure 3. The known-clock PDOP was the same as that in the unconstrained case, and a trend similar to that shown in Figure 2 was observed between satellite geometry and positioning performance. However, the 3D positioning error was reduced by a factor on the order of $10^3$ relative to the unconstrained case. This improvement occurred because the altitude constraint effectively constrained the radial component of the positioning error, as will be discussed in Section 4.2.1. Because the GNSS satellites were clustered within a small angular region as viewed from the lunar-surface receiver, the radial component dominated the 3D positioning error. The altitude constraint effectively limited the radial error, thereby reducing the total 3D positioning error. In this case, the 3D positioning error decreased from approximately 800 m at the beginning to 4 m after four minutes of accumulation, at which point 744 measurements had been accumulated and 251 state variables were estimated.

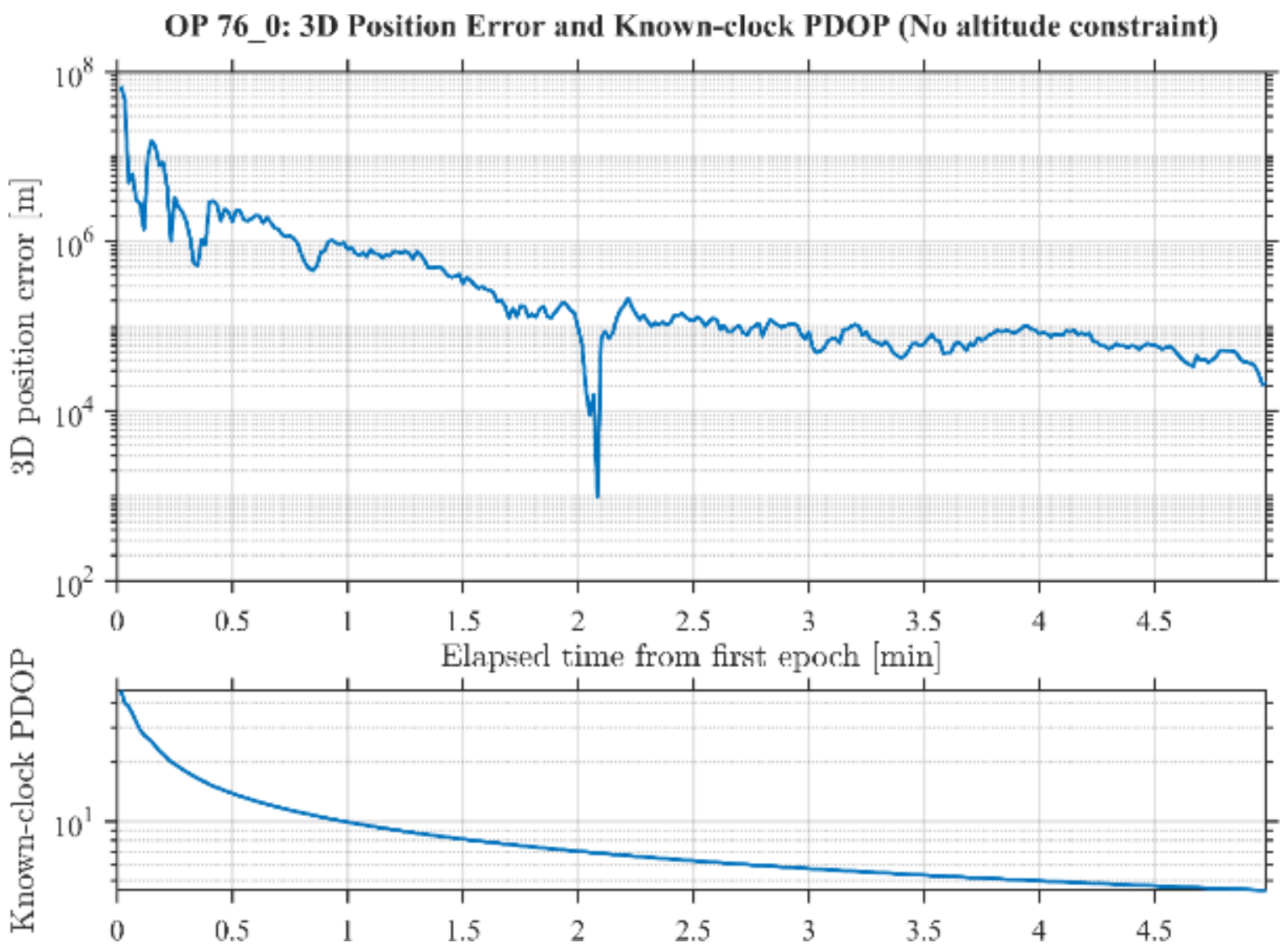

**Figure 2 3D positioning error (m) and the PDOP under the assumption of the perfect knowledge of receiver clock bias**

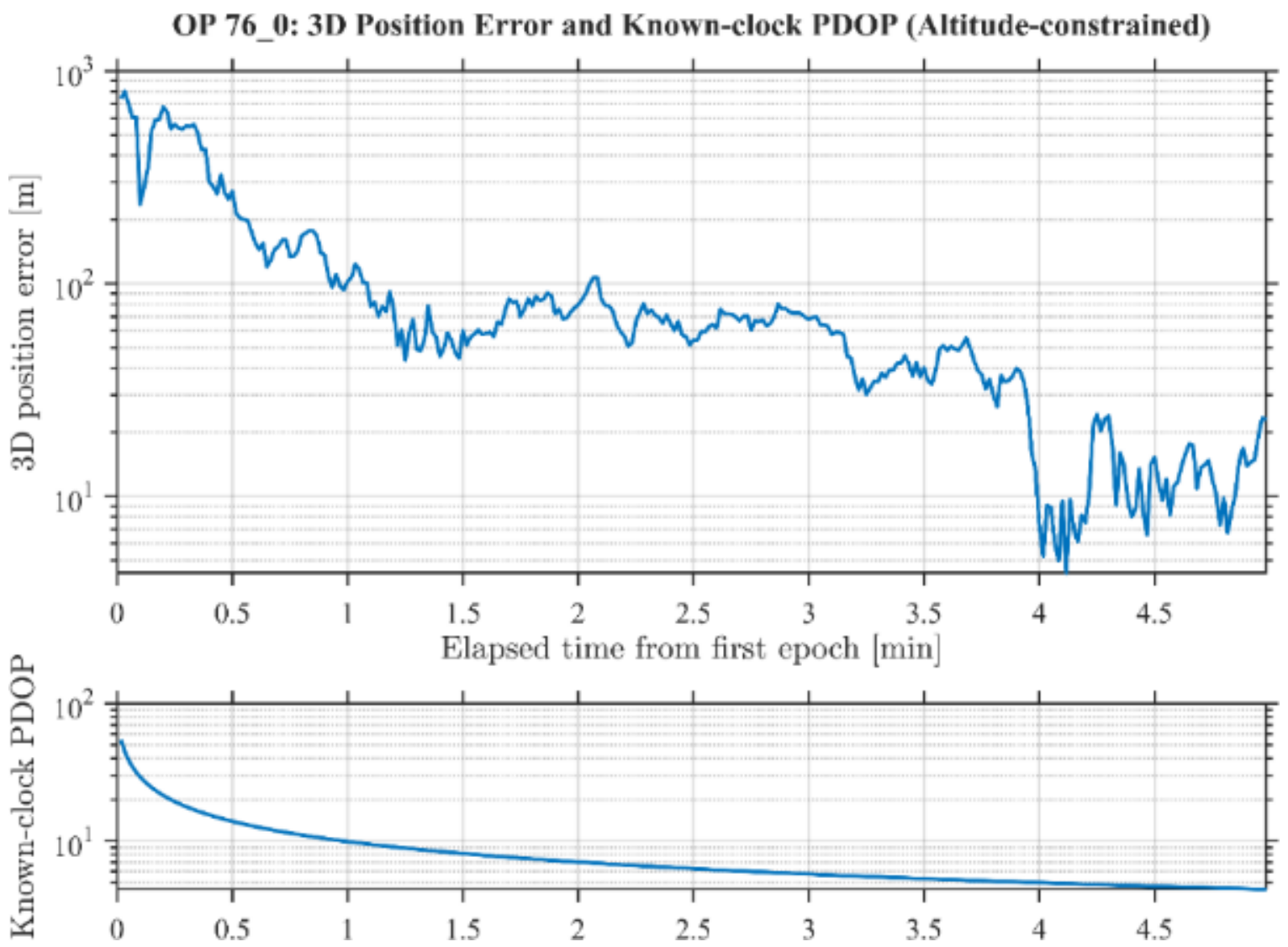

**Figure 3 3D positioning error (m) and the PDOP under the assumption of the perfect knowledge of receiver clock bias when altitude constraint is executed**

A few exceptions to this trend were also observed, in which accumulating additional measurements instead increased the 3D positioning error. These cases shared a common feature: some of the included measurements were of questionable quality. For instance, in OP 38_0, as shown in Figure 4, approximately 0.65 min after the start of accumulation, the straight-line tangential altitude (SLTA) reached −20 km and the carrier-to-noise-density ratio (C/N0) dropped to approximately 20 dB-Hz. Once measurements from this epoch onward were included in the positioning solution, the 3D positioning error began to increase, as shown in Figure 4(a). This behavior is consistent with the signal propagating through the lower terrestrial atmosphere and approaching occultation by the Earth, thereby introducing a bias induced by the crossing of the ionosphere.

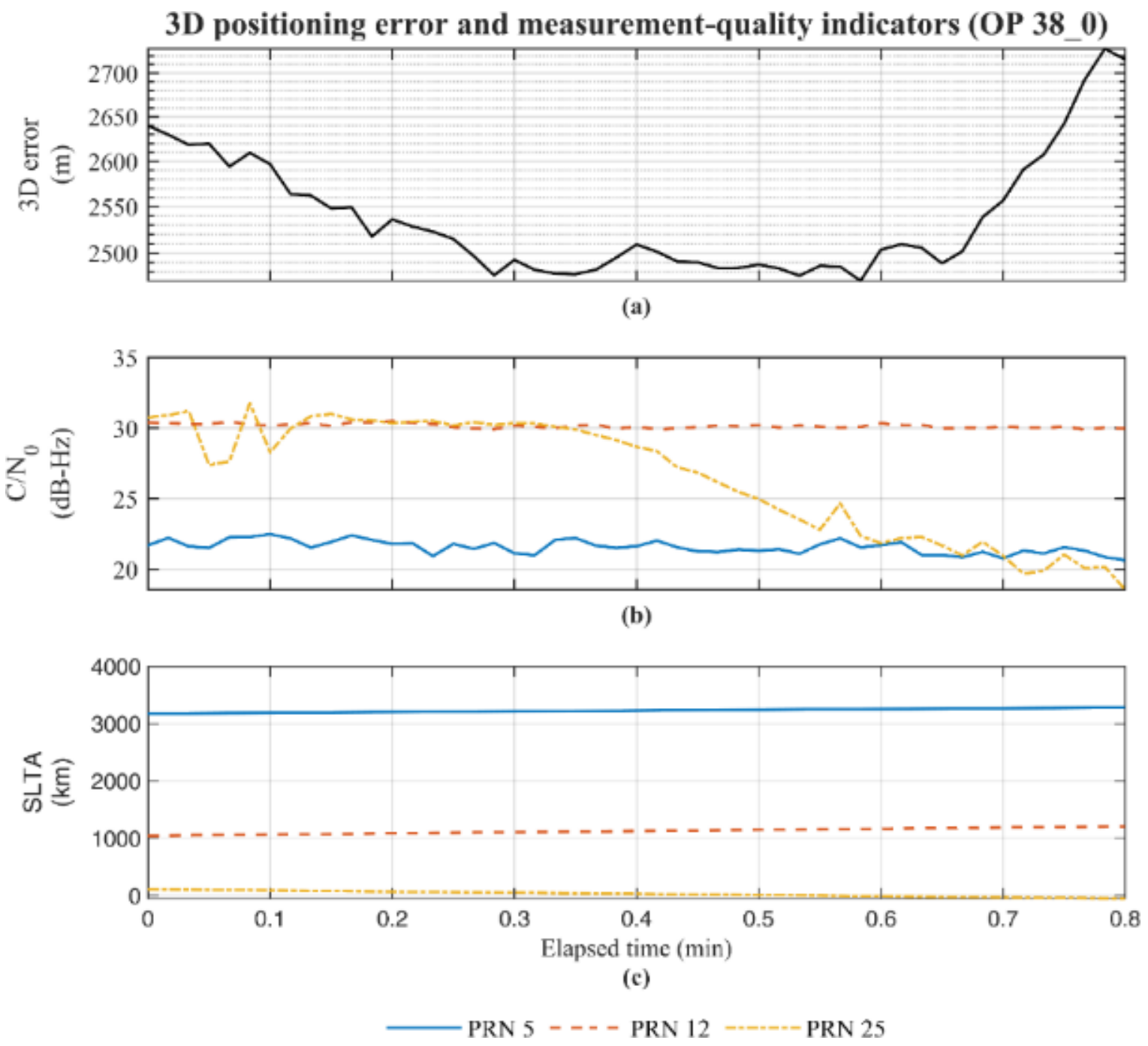

**Figure 4 Evolution of the 3D positioning error as GPS L1 C/A pseudorange measurements were accumulated over a selected interval of OP 38_0: (a) 3D positioning error, (b) carrier-to-noise-density ratio (C/N0), and (c) straight-line tangential altitude (SLTA) of the measurements from PRNs 5, 12, and 25**

Another notable example occurred in OP 40_0, as shown in Figure 5. The C/N0 of PRN 5 began to fluctuate approximately 0.3 min after the start of accumulation, at which point the positioning error began to increase. Given that the SLTA was approximately 80 km at that epoch, the fluctuations may have been associated with atmospheric propagation effects. The 3D positioning error continued to increase for approximately 2.5 min after the onset of the C/N0 fluctuation, after which the error trend turned downward as higher-quality measurements were accumulated and outweighed the influence of the lower-quality measurements collected around t=0.3 min.

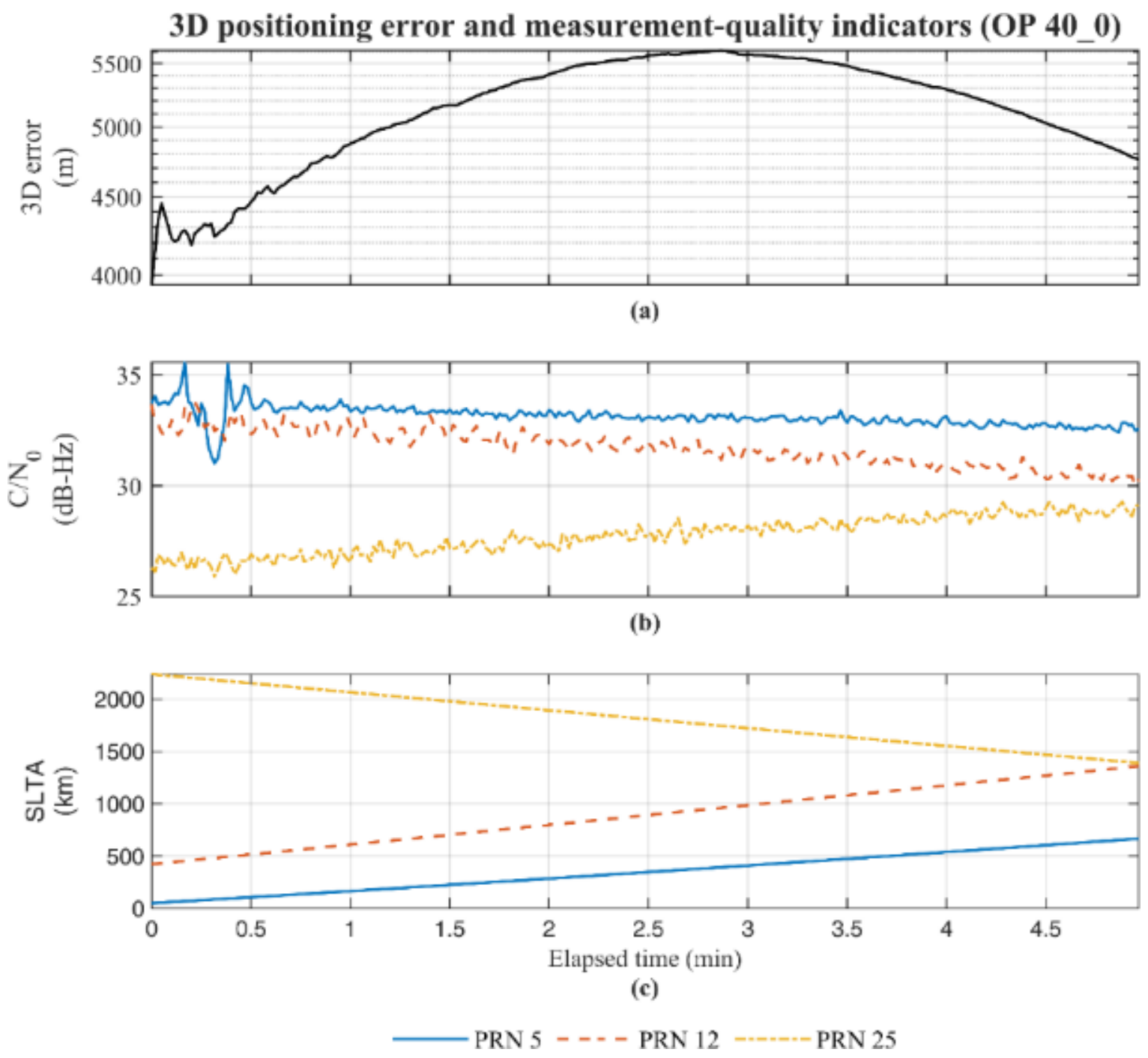

**Figure 5 Evolution of the 3D positioning error as GPS L1 C/A pseudorange measurements were accumulated over a selected interval of OP 40_0: (a) 3D positioning error, (b) carrier-to-noise-density ratio (C/N0), and (c) straight-line tangential altitude (SLTA) of the measurements from PRNs 5, 12, and 25**

When a minimum SLTA threshold of 500 km was applied and accumulation was initiated after the SLTA of PRN 5 exceeded the threshold, a decreasing trend in the 3D positioning error was observed as more measurements were accumulated, as shown in Figure 6. Applying the SLTA threshold excluded the initial fluctuation observed in Figure 5. These results are consistent with the previous simulation study conducted by Iiyama & Gao (2025), in which ionospheric and plasmaspheric delays were predicted to become more significant at lower tangential altitudes.

Another notable aspect of single-OP multi-epoch positioning is the number of measurements available per epoch. Four OPs (OP 38_0, 40_0, 76_0, and 77_1) contained epochs with up to three measurements, whereas the other three OPs (OP 74_0, 77_0, and 78_1) had at most two measurements at any epoch. Figure 7 shows the 3D positioning-error trends over time for different OPs conducted around March 14. OP 76_0 and OP 77_1, denoted by the red and purple lines, respectively, had up to three measurements per epoch, whereas the other OPs, denoted by the blue, green, and yellow lines, had at most two. The 3D positioning-error levels differed clearly between the three-measurement and two-measurement OP groups. For the three-measurement group, the 3D positioning error decreased below 100 m within one minute. However, for the two-measurement group, the 3D positioning errors remained greater than 1 km for most of the five-minute accumulation window.

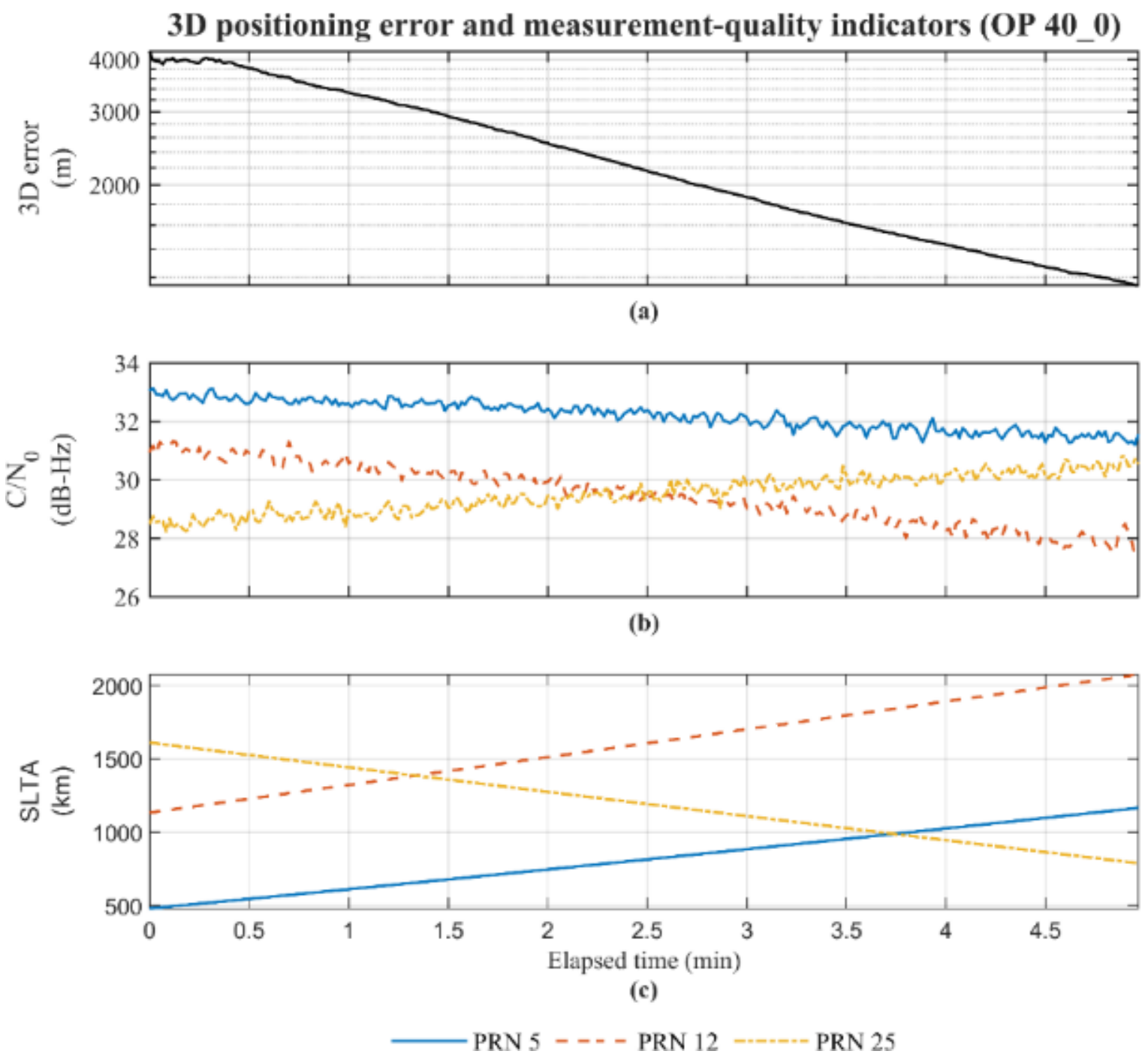

**Figure 6 Evolution of the 3D positioning error as GPS L1 C/A pseudorange measurements satisfying the SLTA threshold were accumulated over a selected interval of OP 40_0: (a) 3D positioning error, (b) carrier-to-noise-density ratio (C/N0), and (c) straight-line tangential altitude (SLTA) of the measurements from PRNs 5, 12, and 25**

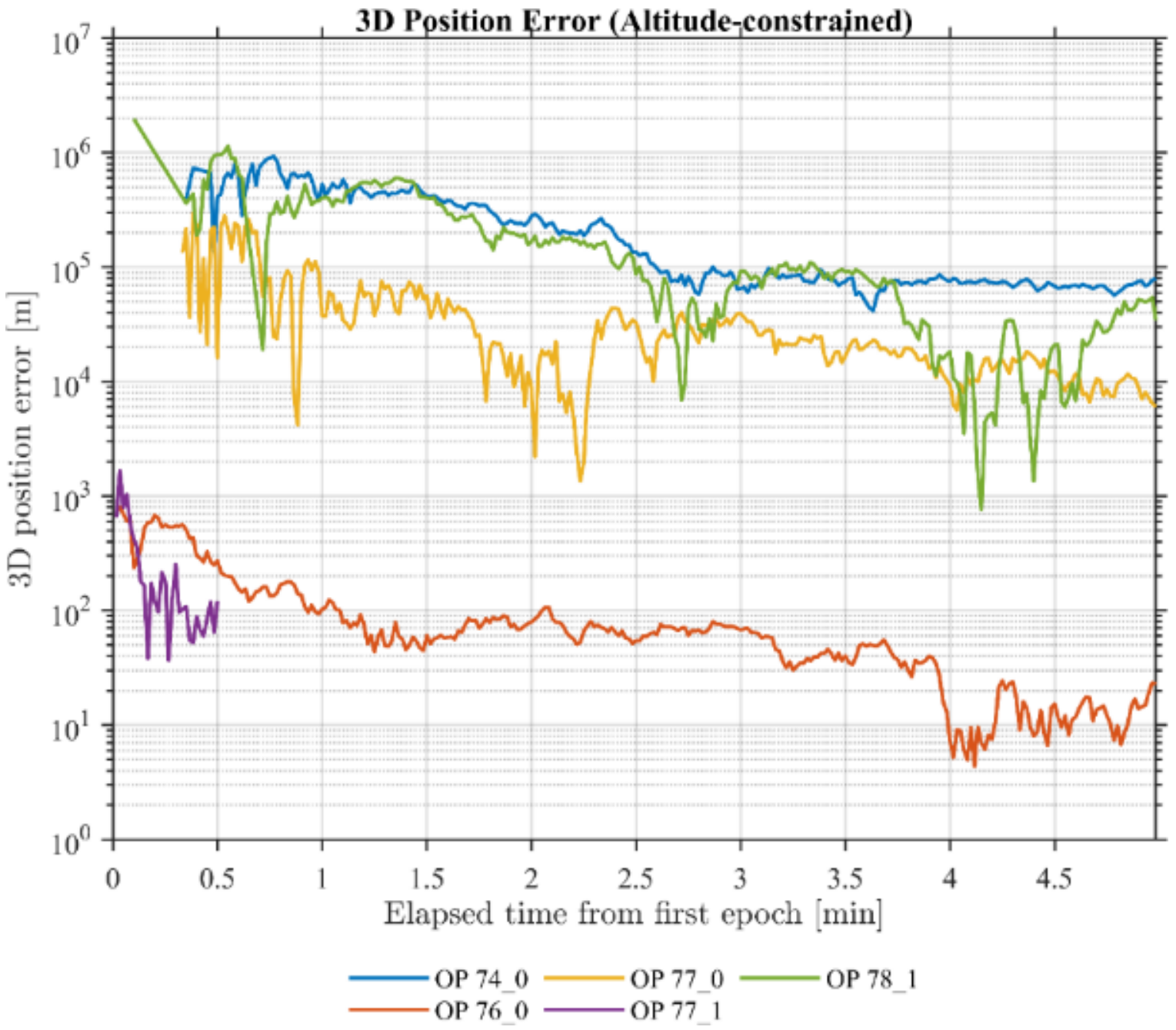

**Figure 7 Evolution of the altitude-constrained 3D positioning error with sequential accumulation of GPS L1 C/A pseudorange measurements within selected lunar-surface operations.**

### 4.2. Randomized epoch-selection results

The single-OP multi-epoch positioning results in Section 4.1 indicate that positioning performance on the lunar surface is affected by the altitude constraint, the number of accumulated epochs, the number of measurements available per epoch, and the straight-line tangential altitude (SLTA). Although positioning performance improved with the application of the altitude constraint, the accumulation of a larger number of measurements, and the exclusion of low-SLTA measurements, these results may not represent lunar-surface positioning performance under all operating conditions. To provide a more general statistical assessment, random epoch-selection tests were conducted by using epochs randomly selected from the publicly available LuGRE TLM_RAW datasets under different test settings, as described in Table 1. These tests were used to characterize the resulting positioning performance and to evaluate the effects of the factors identified from the single-OP multi-epoch positioning tests.

| Section | Altitude constraint | SLTA threshold | Minimum measurements per epoch | Epochs per OP |
|---|---|---|---|---|
| 4.2.1. | Varied | $-\infty$ | OP-specific maximum | 1 |
| 4.2.2. | $\sigma_h = 1\sigma_\rho$ | Varied | 3 (selected OPs) | 30 |
| 4.2.3. | $\sigma_h = 1\sigma_\rho$ | $-\infty$ | Varied | 10 |
| 4.2.4. | $\sigma_h = 1\sigma_\rho$ | $-\infty$ | 2 | Varied |

**Table 1 Random epoch-selection test configurations used to evaluate the effects of the altitude constraint, SLTA threshold, minimum number of measurements per epoch, and number of accumulated epochs per operation**

#### 4.2.1. Effects of Altitude Constraint

As briefly mentioned in Section 4.1, imposing an altitude constraint significantly improved lunar-surface positioning accuracy. From an operational perspective, such an altitude prior is realistic for a lunar-surface receiver because the receiver is known to be located on or near the lunar surface, and approximate altitude information can be obtained from knowledge of the landing site or lunar topographic models.

To evaluate this effect, the random epoch-selection test results are presented in this subsection. In this test, one epoch was randomly selected from each OP for accumulation, resulting in the accumulation of 18 measurements from seven different epochs. Since the first and last OPs used in the LuGRE lunar-surface dataset were conducted on March 3 and March 16, respectively, the accumulated epochs could span nearly two weeks. Figure 8 illustrates the impact of the altitude constraint on the 3D positioning error. In the random epoch-selection tests, the 67th percentile of the 3D positioning error was 5.8 km without the altitude constraint, whereas it was reduced to 970 m when the altitude constraint was applied. The altitude constraint reduced the positioning error by approximately 83%, with the most pronounced improvement occurring in the radial direction, defined here as the direction from the true LuGRE receiver position on the lunar surface toward the Earth center. The histogram in Figure 9 shows that, without the altitude constraint, the radial errors were widely distributed and reached up to 14 km. In contrast, applying the altitude constraint confined most radial errors within 3 km. This result can be attributed to the unfavorable GNSS geometry at lunar distances, where the GNSS satellites are clustered within a small region of the sky as viewed from the LuGRE receiver. Under such geometry, GNSS pseudorange measurements alone provide only a weak constraint in the radial direction. The altitude constraint therefore primarily reduces the 3D positioning error by constraining the radial component.

Additional random epoch-selection tests were conducted to evaluate a scenario in which the a priori altitude information had a larger uncertainty. In this case, the altitude pseudo-measurement was assigned a lower weight than the pseudorange measurements in the weighted least-squares estimation. When the altitude-constraint uncertainty ($\sigma_h$) was set to 35 times the pseudorange measurement uncertainty ($\sigma_\rho$), the positioning performance degraded, with the 67th and 95th percentiles increasing to 1.1 km and 1.9 km, respectively. This trend is consistent with the aforementioned results, which showed that the altitude constraint significantly improves positioning performance.

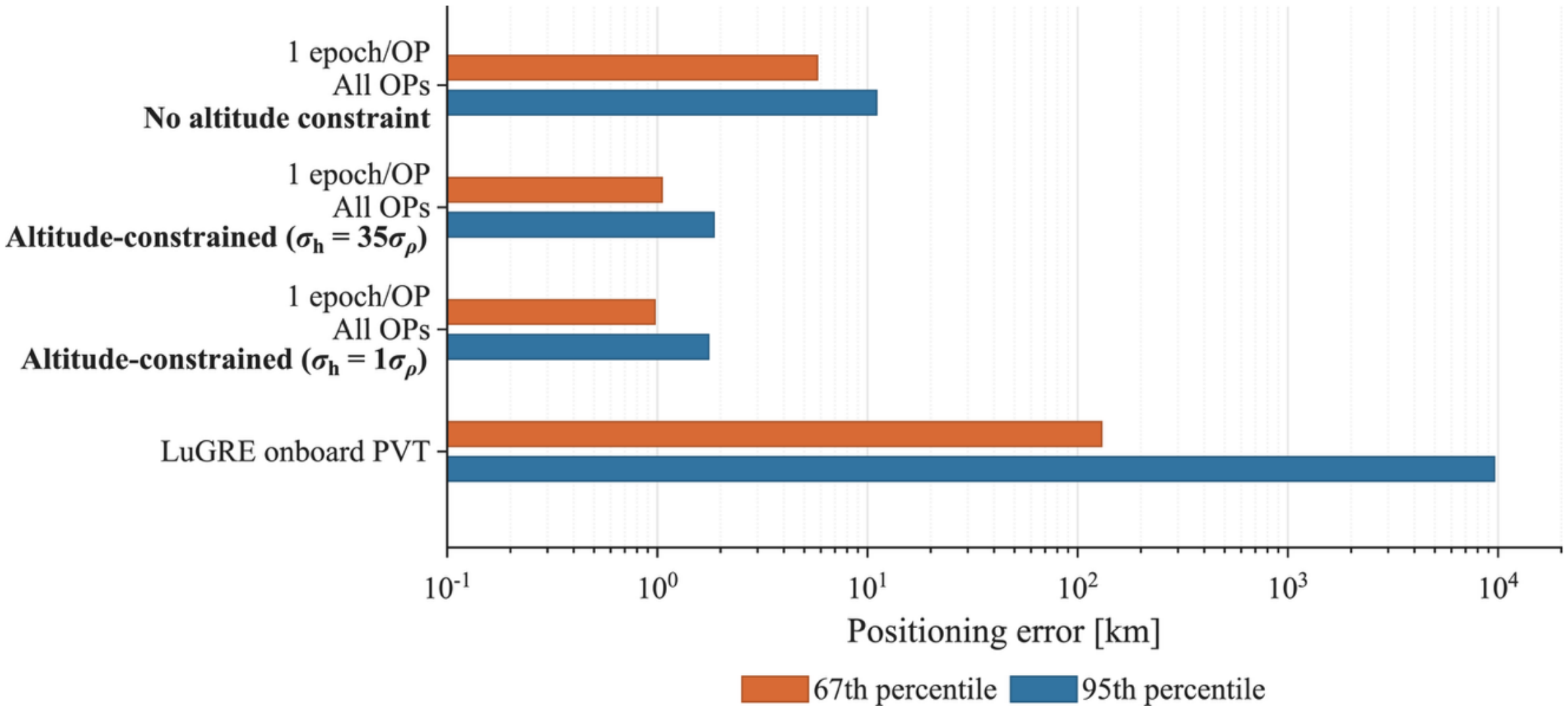

**Figure 8 Random epoch-selection 3D positioning-error percentiles obtained using one randomly selected epoch from each lunar-surface operation under different altitude-constraint settings. The 67th and 95th percentiles are shown for the unconstrained case and for altitude constraints with $\sigma_h/\sigma_\rho = 35$ and $\sigma_h = \sigma_\rho$. For comparison, the 67th and 95th percentiles of the LuGRE onboard PVT positioning errors, calculated after coordinate transformation of the positions provided in the TLM_NAV dataset, are also shown.**

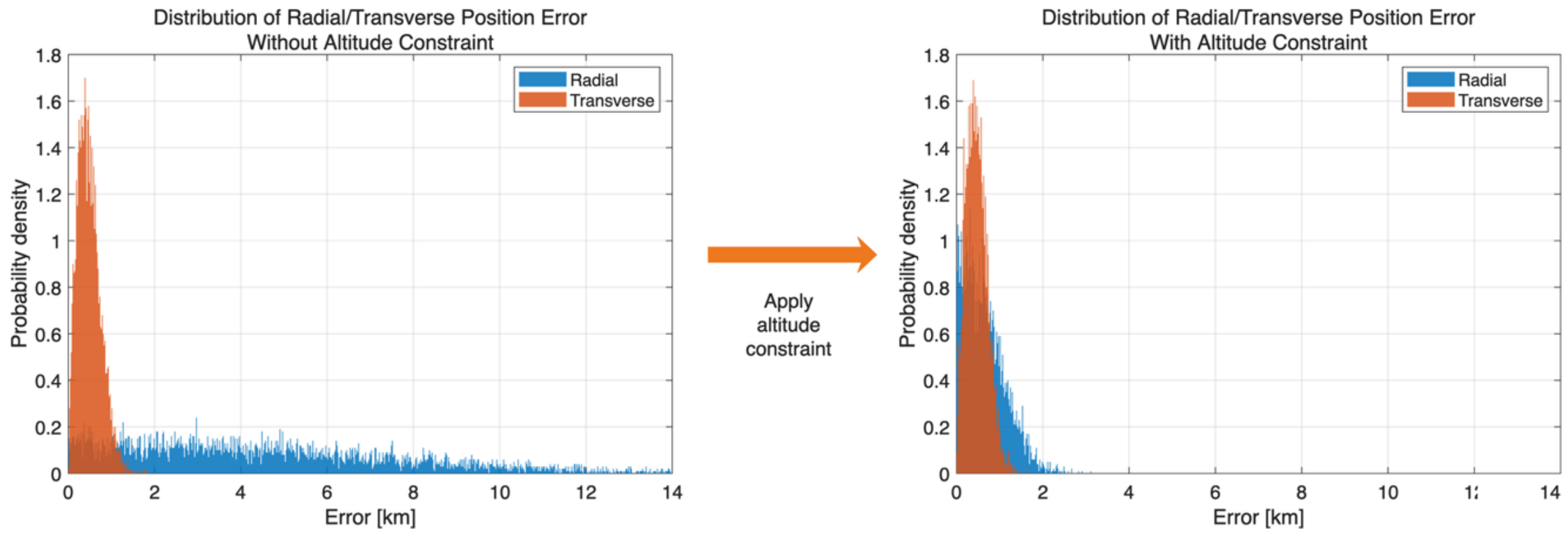

**Figure 9 Probability-density distributions of the radial and transverse positioning errors for (a) positioning without an altitude constraint and (b) positioning with an altitude constraint.**

### 4.2.2. Effects of the SLTA threshold

This subsection examines how excluding low-SLTA measurements from the positioning process affects positioning performance. According to Iiyama & Gao (2025), GNSS signal reception in the vicinity of the Moon requires consideration of ionospheric and plasmaspheric effects on signal propagation and measurement quality. Iiyama & Gao (2025) reported that low-SLTA signals experienced larger group delays than high-SLTA signals. Following their study, the SLTA thresholds considered in this analysis were set to 0, 500, and 1,000 km. Figure 10 shows the impact of the SLTA threshold on accumulated pseudorange-based lunar-surface positioning performance. To isolate the effect of excluding low-SLTA measurements, all other processing settings, including the number of accumulated epochs, the altitude-constraint uncertainty, and the minimum number of measurements required for epoch selection, were kept unchanged.

As noted in Section 4.1, applying a higher SLTA threshold improved positioning performance in the single-OP case. This finding is further supported by the random epoch-selection tests, in which the most restrictive SLTA threshold considered in this study, 1,000 km, yielded the most accurate position estimates. Quantitatively, the 1,000 km SLTA threshold reduced the 67th percentile of the positioning error to 110 m. In contrast, when no SLTA threshold was applied, the 67th percentile was approximately three times as large as that obtained with the 1,000 km threshold.

However, applying a higher SLTA threshold also excluded a large fraction of the available epochs, thereby limiting the number of epochs available for accumulation. With the 1,000 km SLTA threshold, no epochs with three measurements remained available. In contrast, under the 500 km SLTA threshold, the OPs whose maximum per-epoch measurement count was three still provided epochs containing three measurements. Therefore, although increasing the SLTA threshold from 500 to 1,000 km prevented the positioning process from benefiting from the larger number of measurements available per epoch, excluding measurements more strongly affected by terrestrial ionospheric and plasmaspheric propagation effects provided a greater overall benefit than retaining the additional measurements. This result indicates an inherent trade-off between measurement quality and measurement availability, which should be carefully considered when designing lunar-surface positioning strategies.

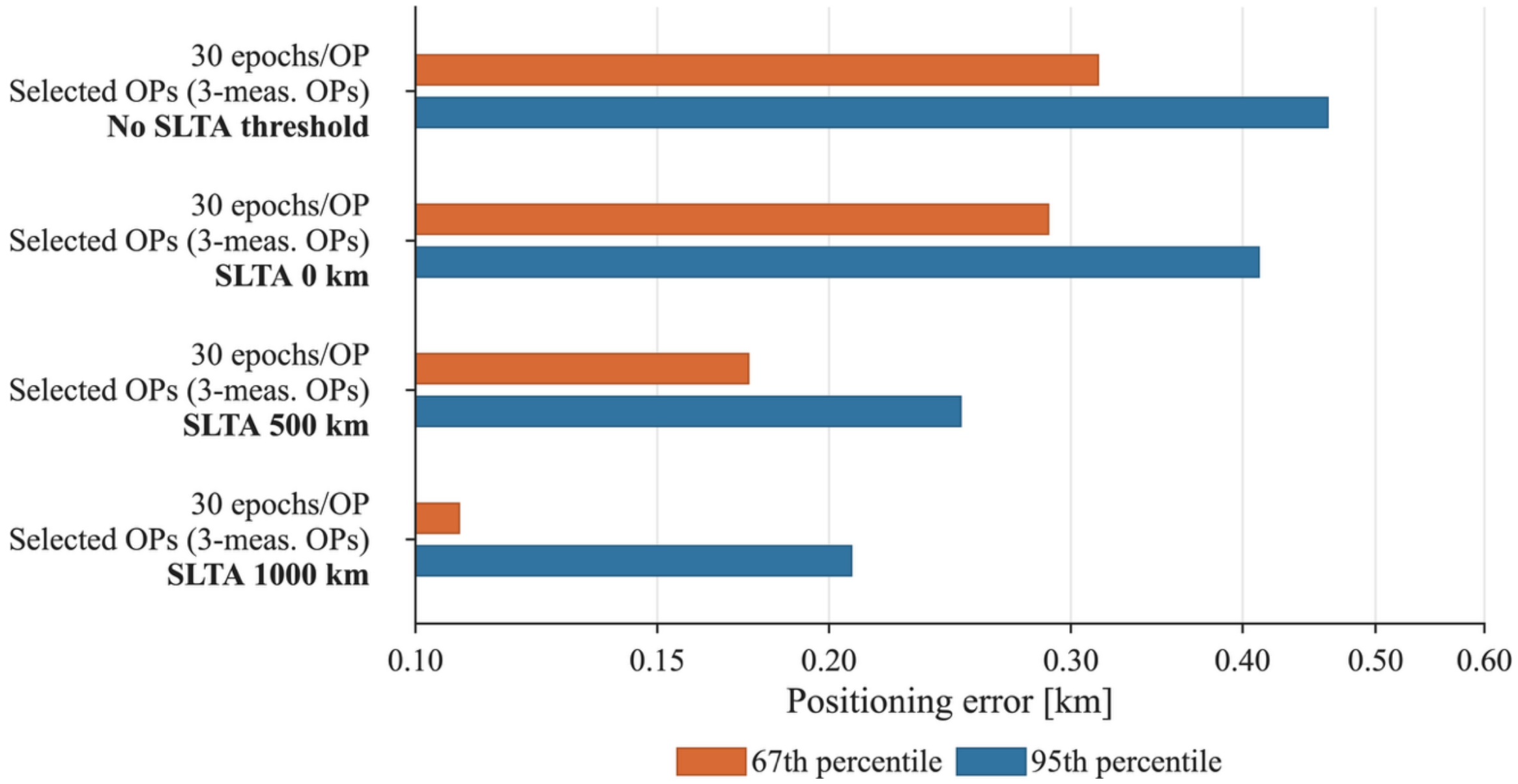

**Figure 10 Effects of the straight-line tangential altitude (SLTA) threshold on the random epoch-selection 3D positioning-error percentiles. Thirty epochs were randomly selected from each operation containing epochs with at least three measurements.**

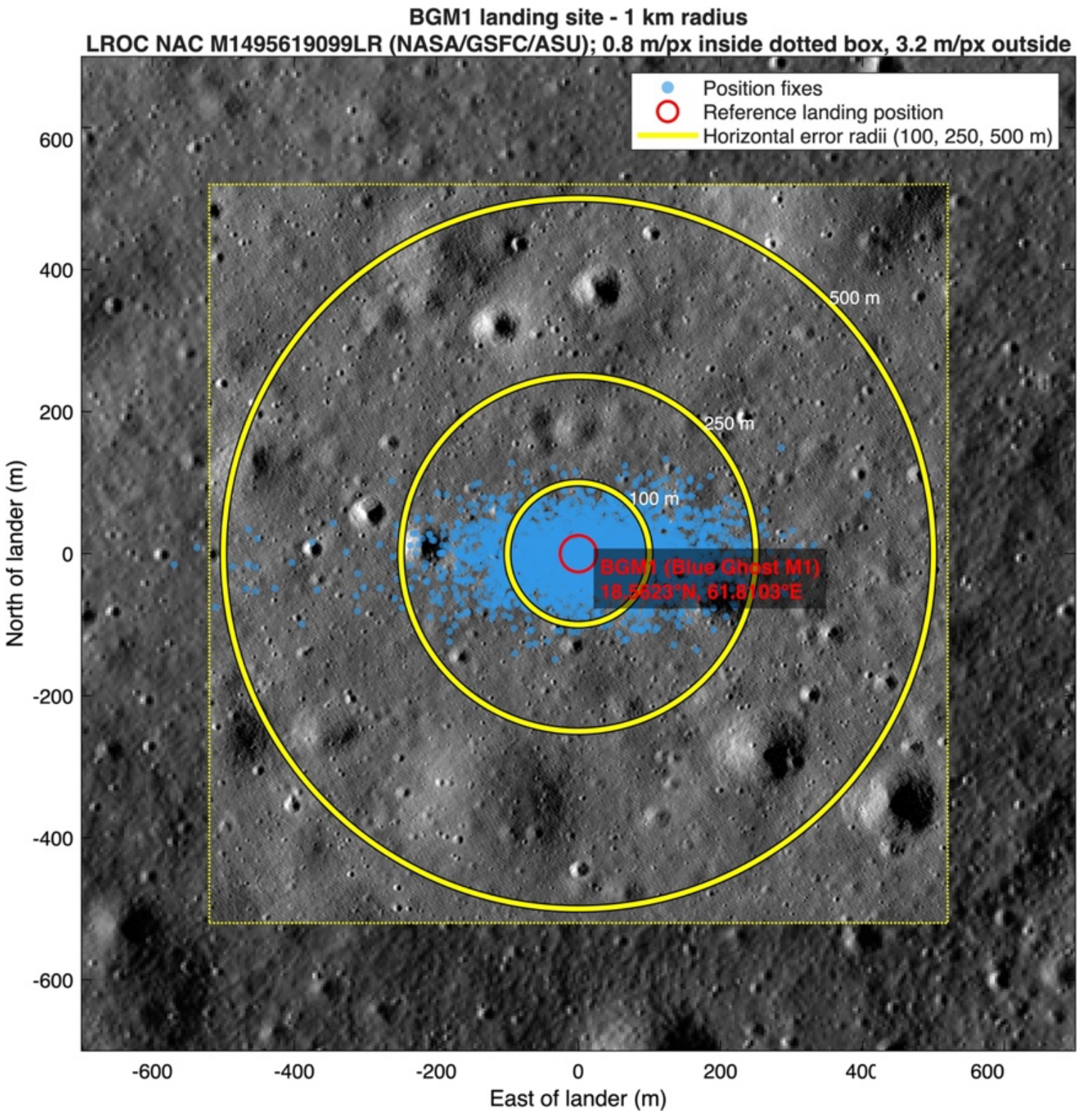

**Figure 11 Position estimates from randomized multi-epoch batch positioning overlaid on imagery of the BGM1 landing site (Robinson, 2025b). The estimates were obtained using an altitude constraint, an SLTA threshold of 1,000 km, and at least two pseudorange measurements per epoch.**

#### 4.2.3. Effects of the minimum number of measurements from each epoch

This subsection examines the impact of the minimum number of measurements required for epoch selection on positioning performance. As discussed in Section 4.1, the number of measurements available at the selected epochs is closely related to the performance of single-OP multi-epoch positioning using accumulated pseudoranges. This finding was further evaluated through random epoch-selection tests, in which 10 epochs were selected from each lunar-surface OP and an a-priori altitude constraint was incorporated together with the pseudorange measurements. Depending on the selection criterion, epochs were randomly selected from either all seven lunar-surface OPs or only the OPs satisfying the required minimum measurement count. Figure 12 summarizes the random epoch-selection test results for different minimum measurement-count thresholds applied during epoch selection.

When epochs with at least one measurement were allowed to be selected, the 95th-percentile 3D positioning error reached 1.9 km, with a 67th-percentile error of 800 m. When only epochs with at least two measurements were randomly selected, the 95th percentile decreased to 650 m, while the 67th percentile decreased to 320 m. This corresponds to an approximately 60% reduction in both percentiles. When the selected epochs were further restricted to those having the maximum number of measurements available within each OP, the 95th percentile decreased slightly further, from 650 m to 600 m, corresponding to an additional 8% reduction. In contrast, the 67th percentile increased by approximately 10%. Nevertheless, the minimum and maximum errors were reduced from 2.2 m to 2.1 m and from 1.6 km to 1.0 km, respectively.

These results indicate that requiring a larger minimum number of measurements per selected epoch generally improves positioning performance. This behavior can be explained by the structure of the accumulated measurement model. In the multi-epoch formulation, the receiver position is common across the accumulated epochs, whereas a separate receiver clock bias is estimated for each epoch. Therefore, adding an epoch with only one pseudorange measurement introduces one additional observation but also one additional receiver-bias state. Such an epoch provides little independent geometric information for estimating the common receiver position. In contrast, an epoch with multiple measurements contributes more observations while still introducing only one epoch-wise receiver bias, thereby improving the observation-to-unknown ratio and generally leading to better positioning accuracy.

The slight increase in the 67th percentile for the maximum-measurement case suggests that measurement count alone does not fully determine positioning performance; the geometry, C/N0, and SLTA of the selected measurements can also affect the solution quality. When epoch selection was further restricted to epochs containing at least three measurements, positioning performance degraded because three of the seven lunar-surface OPs were excluded. These OPs contained at most two measurements at any epoch and therefore did not contribute to the accumulated pseudorange set under this selection criterion. Consequently, the loss of temporal and geometric diversity resulting from the exclusion of these OPs outweighed the benefit of retaining only epochs with larger measurement counts.

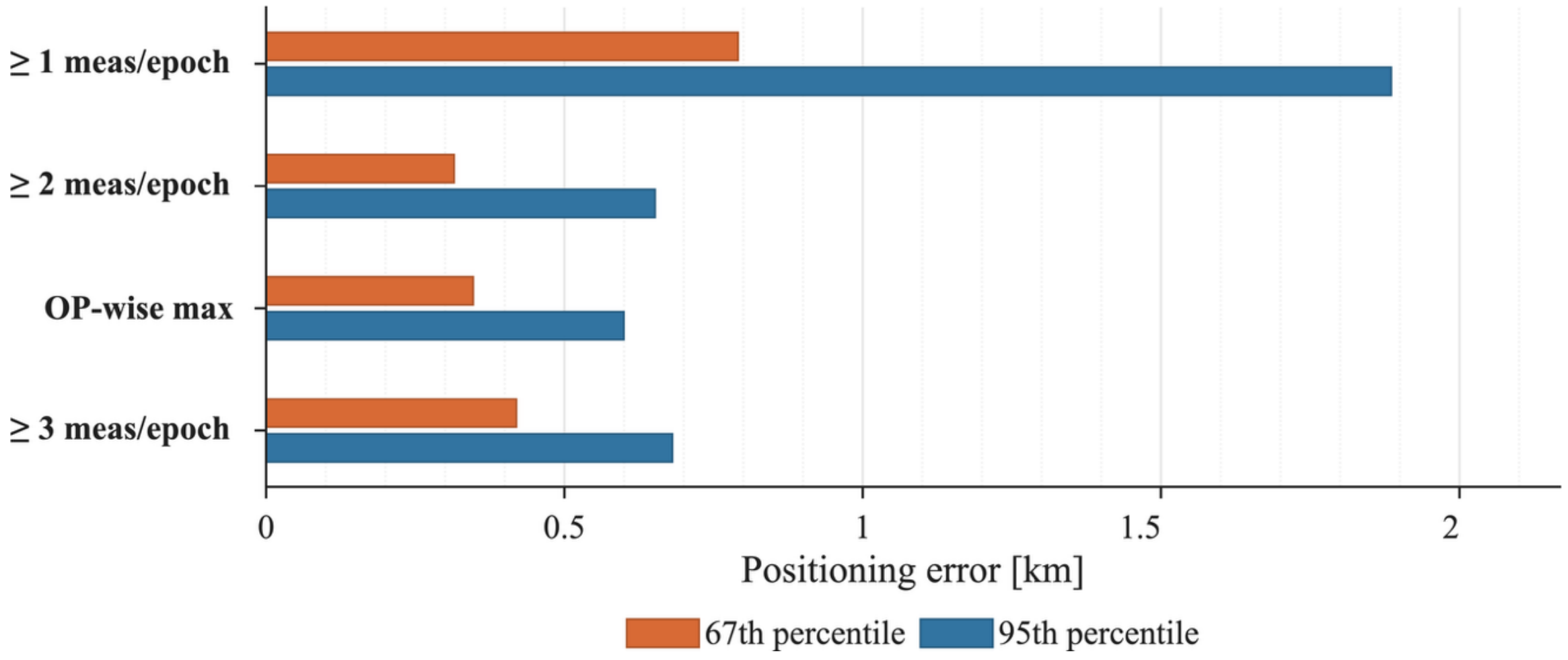

**Figure 12 Random epoch-selection 3D positioning-error percentiles obtained using different epoch-selection criteria. The 67th and 95th percentiles are compared for randomly selected epochs containing at least one measurement, randomly selected epochs containing at least two measurements, and epochs with the maximum number of available measurements within each operation.**

#### 4.2.4. Effects of the number of epochs accumulated

The number of epochs accumulated for estimating the position of a static receiver on the lunar surface was shown in Section 4.1 to play a key role in determining positioning performance. Because multi-epoch accumulation is central to the proposed approach, its impact was further evaluated through random epoch-selection tests. Since the number of epochs with the maximum measurement count in each OP was limited, the OPs were divided into two groups based on their operation times, and epochs were randomly selected from each group. Specifically, OPs 38_0 and 40_0 were grouped together, while the remaining five OPs formed the second group. Epochs with fewer than two measurements were excluded, and the altitude constraint was applied.

The random epoch-selection test results are consistent with the single-OP multi-epoch positioning results described in Section 4.1, which showed that accumulating more epochs generally reduced the 3D positioning error. As more epochs were accumulated, both the 67th and 95th percentiles tended to decrease, although the rate of improvement gradually diminished. For example, the 67th percentile decreased from 260 m with 50 epochs accumulated per group to 130 m with 200 epochs accumulated per group. However, the improvement rate decreased as the number of accumulated epochs increased. When the number of accumulated epochs per group was increased from 50 to 100, from 100 to 150, and from 150 to 200, the corresponding reductions in the 67th percentile were 30%, 18%, and 10%, respectively. This diminishing improvement may be attributed to the progressively smaller increase in geometric diversity provided by additional epochs, together with the increase in the number of state variables resulting from the estimation of a separate receiver clock bias for each accumulated epoch.

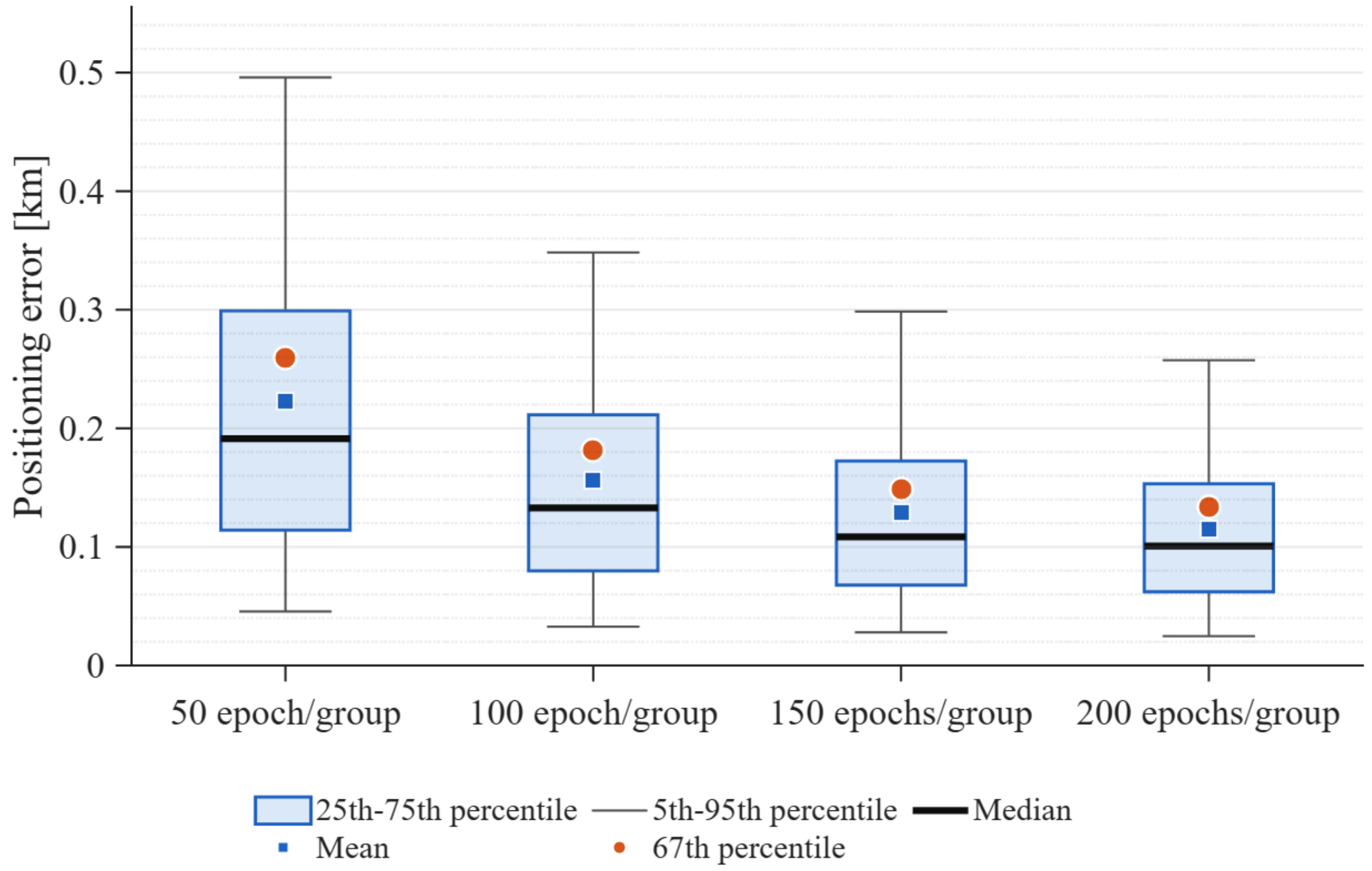

**Figure 13 Random epoch-selection 3D positioning-error distributions for different numbers of accumulated epochs per group. The boxes represent the 25th–75th percentile ranges, the whiskers represent the 5th–95th percentile ranges, and the lines, squares, and circles indicate the median, mean, and 67th percentile, respectively.**

## 5. CONCLUSIONS

This study investigated lunar-surface positioning using GPS L1 C/A pseudorange measurements generated by the LuGRE receiver and accumulated over multiple epochs in a batch least-squares formulation. By exploiting the stationary position of the receiver in the lunar-centered, lunar-fixed frame, pseudorange measurements collected at different times were combined while a separate receiver clock bias was estimated for each accumulated epoch.

Single-operation analyses identified several factors affecting positioning performance, including the use of an altitude constraint, the minimum number of measurements required for epoch selection, the straight-line tangential altitude threshold, and the number of accumulated epochs. With the altitude constraint applied, accumulating epochs containing three pseudorange measurements from OP77_1 achieved 100-m-level positioning accuracy after several tens of seconds of accumulation. Similarly, OP76_0 achieved comparable accuracy after approximately one minute of accumulation.

The effects of these factors were further evaluated through random epoch-selection tests using the LuGRE lunar-surface pseudorange data. The altitude constraint substantially reduced the weakly observable radial error component and consequently reduced the overall 3D positioning error. Excluding measurements below the prescribed SLTA threshold also improved positioning performance. However, increasing the SLTA threshold reduced the number of available measurements, indicating a trade-off between measurement quality and measurement availability. Increasing the number of accumulated epochs generally reduced the 3D positioning error, although the incremental improvement diminished as more epochs were added. The number of measurements available per selected epoch also affected positioning performance, particularly because a separate receiver clock-bias state was estimated for each epoch.

In the randomized epoch-selection experiments across multiple OPs, applying an altitude constraint and an SLTA threshold of 1,000 km, requiring at least two measurements per epoch, and accumulating 120 epochs yielded 67th- and 95th-percentile 3D positioning errors of approximately 110 m and 210 m, respectively. These results were obtained using only single-frequency, single-constellation GPS L1 C/A pseudorange measurements. To the best of the authors' knowledge, this result represents the first demonstration of 100-m-level positioning using actual GNSS measurements collected on the lunar surface. The resulting errors were substantially smaller than the onboard single-epoch positioning errors reported during the LuGRE lunar-surface operations, which had a 67th-percentile value of 130 km. These results demonstrate that multi-epoch pseudorange accumulation can mitigate the limitations imposed by sparse signal availability and unfavorable GNSS geometry on the lunar surface.

Future studies will investigate batch weighted least-squares positioning using multi-constellation GNSS measurements and the combination of measurements from different frequency bands to mitigate ionospheric effects. Given the low carrier-to-noise-density ratios of GNSS signals at lunar distances, future studies will also evaluate the trade-off between ionospheric error mitigation and measurement noise amplification associated with ionosphere-free combinations.